\documentclass[aps,prd,twocolumn,groupedaddress,nofootinbib]{revtex4-1}

\pdfoutput=1

\usepackage{amsfonts,amsmath,amssymb}
\usepackage{mathrsfs}
\usepackage{nicefrac}
\usepackage{graphicx}
\graphicspath{{graphics/}} 
\usepackage{color}
\usepackage{soul} 
\usepackage{hyperref}
\usepackage{accents}
\usepackage[normalem]{ulem}
\usepackage{cleveref}
\hypersetup{
    colorlinks=true, 
    linktoc=page,    
    linkcolor=blue,  
    urlcolor=black
}

\usepackage{bookmark}
\bookmarksetup{
  numbered, 
}
\usepackage{lmodern} 
\usepackage[T1]{fontenc}
\usepackage[utf8]{inputenc} 
\usepackage{amssymb,amsmath} 
\usepackage{textcomp}
\usepackage{mathrsfs}
\usepackage{braket}
\usepackage{slashed}
\usepackage{tensor}
\usepackage[subnum]{cases}
\usepackage{mathtools}
\usepackage{bbm}

\newcommand*{\defeq}{\mathrel{\vcenter{\baselineskip0.5ex \lineskiplimit0pt
                     \hbox{\scriptsize.}\hbox{\scriptsize.}}}%
                     =} 
\newcommand{\de}{\partial}
\newcommand{\hsp}{\hspace{1.5 cm}}
\newcommand{\pder}[2]{\frac{\de#1}{\de#2}}

\renewcommand{\O}{\mathcal{O}}
\newcommand{\e}{\epsilon}
\renewcommand{\t}{\theta}
\newcommand{\p}{\phi}

\renewcommand{\a}{\alpha}

\newcommand{\m}{\mu}
\renewcommand{\r}{\rho}
\newcommand{\n}{\nu}
\newcommand{\G}{\mathcal{G}}

\newcommand{\del}{\nabla}

\newcommand{\Lam}{\Lambda}
\renewcommand{\l}{\ell}
\renewcommand{\d}{\mathrm{d}}

 \newcommand{\vc}[1]{\mathbf{#1}}

\newcommand{\h}{\hbar}

\newcommand{\norm}[1]{\left\lVert #1 \right\rVert}
\newcommand{\tld}[1]{\widetilde{#1}}

\begin{document}

\title{Rotating black hole solutions in relativistic analogue gravity}

\author{Luca Giacomelli}
\email[]{lucagiacomelli92@gmail.com}
\affiliation{University of Trento,
Via Sommarive 14, 38123 Povo (TN), Italy}

\author{Stefano Liberati}
\email[]{liberati@sissa.it}

\affiliation{SISSA, 
Via Bonomea 265, 34136 Trieste, Italy and INFN, Sezione di Trieste}
                
\begin{abstract}
Simulation and experimental realization of acoustic black holes in analogue gravity systems have lead to a novel understanding of relevant phenomena such as Hawking radiation or superradiance. We explore here the possibility to use relativistic systems for simulating rotating black hole solutions and possibly get an acoustic analogue of a Kerr black hole. In doing so we demonstrate a precise relation between non-relativistic and relativistic solutions and provide a new class of vortex solutions for relativistic systems. Such solutions might be used in the future as a test bed in numerical simulations as well as concrete experiments.
\end{abstract}

\maketitle

\section{Introduction}
Black holes are among the most peculiar objects predicted by general relativity (GR) and as such provide an arena where interesting effects associated to gravity, both classical and quantum, can become manifest. However, such effects are quite difficult to study in astrophysics as they are faint or masked by more complex physical processes. Of course, the recent detection of gravitational waves~\cite{LIGO2016} opens a new era for observational constraints on black hole models and physical properties but even with such advancements the intensity of some of the predicted effects is too weak to be realistically detectable in the near future.

In the past few decades, starting from a seminal idea by Unruh \cite{unruh81}, a lot of interest has been directed towards analogue models of gravity, that is condensed matter systems that reproduce kinematic aspects of the physics in curved spacetimes, and in particular also of black holes (for an exhaustive review see \cite{blv-review}).
One of the most studied systems is the (non-relativistic) barotropic, inviscid and irrotational fluid for which it can be proven that the linearized (acoustic) perturbations can be described as fields on a curved background, described by an \textit{acoustic metric}. This analogy is applicable also to quantum systems that admit such a hydrodynamical limit, for example in non-relativistic Bose--Einstein condensates, that hence provide a way of testing quantum effects such as the Hawking radiation in conditions where the background physics is well known, and experimentally realizable.

Recently, it was established that a similar analogy also emerges in relativistic fluids \cite{bilic1999,visser-relativistic-acoustic} and relativistic condensates \cite{rbec2010}, and it was shown that in this case the structure of the resulting metric is more general (having a so called disformal form). They hence provide a potentially more versatile framework for simulating interesting spacetimes. It has already been shown that analogues of static black hole spacetimes both in the asymptotically flat and asymptotically (A)dS cases \cite{Hossenfelder:2014gwa,Hossenfelder:2015pza,dey2016ads} can be obtained in relativistic condensates, and also cosmological metrics and structures of alternative theories of gravitation can be studied~\cite{rbec2016}. The main scope of this article is to understand if also rotating black holes spacetimes can fit in relativistic analogues and to study their relationship with the non-relativistic solutions. 

In Section \ref{sec:acousticmetric} we will review the emergence of the acoustic metric in a relativistic perfect fluid, the non relativistic limit and the realization of the analogy in relativistic Bose--Einstein condensates. In Section \ref{sec:rel-vs-nonrel} we are going to show a general transformation that implies that, in the case of constant speed of sound, the relativistic acoustic metric is a more general kind of stationary metric than the non-relativistic one (in the sense that every non-relativistic acoustic metric can be brought into a relativistic one) and we are going to use this transformation to obtain the relativistic acoustic analogue of the Schwarzschild spacetime. 
In Section \ref{sec:rotatingacoustic} we are going to show that rotating black holes can be obtained in relativistic analogue gravity, in Section \ref{sec:vortex} we will present (2+1)-dimensional rotating black holes in relativistic fluids and condensates that are relativistic generalizations of the well known non-relativistic \textit{vortex geometry} and in Section \ref{sec:3vortex} we will present two (3+1)-dimensional rotating black holes configurations. Finally, in Section \ref{sec:btz} we will present an acoustic analogue of the rotating BTZ metric, both in non-relativistic and in relativistic systems and in Section \ref{sec:kerrschild} we will discuss the relationship between the relativistic acoustic metric and the Kerr--Schild form of a metric, treating explicitly the Schwarzschild case, with the hope to give some insight towards the simulation of the full Kerr metric.


\section{The acoustic metric}
\label{sec:acousticmetric}
The analogy between the behavior of acoustic perturbances in fluid systems and the propagation of fields in curved spacetimes is expressed by the emergence of a metric describing a curved background. In the case of a relativistic perfect fluid this metric was derived in \cite{bilic1999} and, in a more detailed way, in \cite{visser-relativistic-acoustic}. We review here the basic steps of these derivations for completeness and for setting up our notation.

A relativistic perfect fluid in a spacetime with metric $g_{\m\n}$ is described by a unit timelike 4-velocity field ${V^\m=v^\m/c}$ and has energy-momentum tensor
\begin{equation}
	T_{\m\n}=(e+p)V_\m V_\n+pg_{\m\n},
\end{equation}
where $e$ is the energy density and $p$ is the pressure. We will take the fluid to be {barotropic}, that is with the energy density a function of the pressure alone, and with an {irrotational} flow, condition that is expressed as
\begin{equation}\label{eq:rel-irrotationality}
	v\wedge \d v=0,
\end{equation}
where $v$ is the one-form associated to the 4-velocity field. This implies that we can write the 4-velocity field in terms of a velocity potential $\theta$ as 
\begin{equation}
  V^\m=\frac{g^{\m\n}\del_\n\theta}{\sqrt{-g^{\m\n}\del_\m\theta\del_\n\theta}}.
\end{equation}

The dynamical equations of the fluid are given by the conservation of the energy-momentum tensor
\begin{equation}
	\del_\m \tensor{T}{^\m_\n}=0,
\end{equation}
whose projections give the continuity equation~\cite{andersson-lrr-relativistic-fluid,visser-relativistic-acoustic}
\begin{equation}\label{eq:rel-continuity}
	\del_\m\left(n V^\m\right)=0,	
\end{equation}
where $n$ is the number density, and the Bernoulli equation
\begin{equation}\label{eq:bernoulli-equation}
	\norm{\del \theta}=\frac{[e(p)+p]n_{(p=0)}}{n(p)e_{(p=0)}}.
\end{equation}

The idea is now to consider an expansion for the quantities defining the fluid up to the first order in some small parameter $\e$
\begin{subnumcases}{}
	\theta=\theta_0+\e\theta_1+\dots\Longrightarrow V=V_0+\e V_1+\dots\\
	e=e_0+\e e_1+\dots\\
	p=p_0+\e p_1+\dots
\end{subnumcases}
and define the acoustic perturbations to be the linear terms, while the 0$^{th}$-order quantities are the background.
Inserting these expansions in the fluid equations and linearizing them one obtains that the perturbation of the velocity potential $\theta_1$ satisfies
\begin{equation}
  \del_\m\left\{\frac{n_0}{w_0c_s}\left[g^{\m\n}+\left(1-\frac{c^2}{c_s^2}\right)V_0^\m V_0^\n\right]\del_\n\theta_1\right\}=0,
\end{equation}
where $w_0$ is the background specific enthalpy and ${c_s^2\defeq c^2\left.\frac{\d p}{\d e}\right|_{e_0}}$ is the speed of sound. This equation has the form of a massless Klein--Gordon equation in a curved background (characterized by a metric $\G_{\m\n}$)
\begin{equation}
  \Box\theta_1= \frac{1}{\sqrt{|\G|}}\de_\m \left(\sqrt{|\G|}\G^{\m\n}\de_\n\theta_1\right)=0
\end{equation}
if we identify\footnote{Notice that this equation differs by a factor $\sqrt{|g|}$ from equation (81) of \cite{visser-relativistic-acoustic}, the latter is correct in flat space and with Cartesian coordinates but acquires the extra factor otherwise. This slightly affects equations (82) and (83), but does not propagate any further.}
\begin{equation}
  \sqrt{|\G|}\G^{\m\n}=\sqrt{|g|}\frac{n_0}{w_0c_s}\left[g^{\m\n}+\left(1-\frac{c^2}{c_s^2}\right)V_0^\m V_0^\n\right].
\end{equation}
From this we obtain the relativistic acoustic metric (in $d+1$ dimensions)
\begin{equation}\label{eq:relativistic-acoustic-metric}
  \G_{\m\n}=\left(\frac{n_0}{w_0c_s}\right)^{2/(d-1)}\left[g_{\m\n}+\left(1-\frac{c_s^2}{c^2}\right)[V_0]_\m [V_0]_\n\right],
\end{equation}
where one usually defines
\begin{equation}
  \xi\defeq 1-\frac{c_s^2}{c^2}.
\end{equation}
We will call this kind of metric a \textit{metric of the Gordon form}, since it has the same form of the metric derived by Gordon for the description of the propagation of light in a moving dielectric \cite{gordon1923}. We will actually be interested in the case in which the fluid moves in a flat spacetime, so that $g_{\m\n}=\eta_{\m\n}$.

As is explained in \cite{visser-relativistic-acoustic}, in the non-relativistic limit in which $v^i,\; c_s\ll c$ (and also $p_0\ll e_0$), the acoustic metric (\ref{eq:relativistic-acoustic-metric}) reduces (up to a constant factor) to the acoustic metric describing the propagation of acoustic perturbations in a non-relativistic perfect fluid, expressed by the line element
\begin{equation}\label{eq:nonrel-acoustic-metric}
ds^2=\frac{\r_0}{c_s}\left[-c_s^2\d t^2+(\d x^i-v_0^i\d t)\eta_{ij}(\d x^j-v_0^j\d t)\right],
\end{equation}
where $\r_0$ is the background matter density and ${v_0^i\defeq cV_0^i}$.
Notice that these velocity components are the ones with respect to an orthogonal vector basis $\{\de_i\}$, and not an orthonormal one $\{\hat{\mathbf{e}}_i\}$ that is usually used in non-relativistic context. This is important when working with curvilinear coordinates, and the two bases are related by
\begin{equation}
  \hat{\mathbf{e}}_i=\frac{1}{h_i}\de_i,
\end{equation}
where $h_i\defeq\sqrt{\eta_{ii}}$ are the scale factors. In particular if we indicate with a tilde the components with respect to an orthonormal basis we have
\begin{equation}
  \vc{v}=v^i\de_i=\tld{v}_i\hat{\vc{e}}_i \hspace{.5cm} \Longrightarrow\hspace{.5cm} \tld{v}_i=h_iv^i=\frac{v_i}{h_i}\hspace{.5cm} \mbox{(no sum)}.
\end{equation}

\subsection{Relativistic Bose--Einstein condensates}
In non-relativistic analogue gravity there has been a great interest in Bose--Einstein condensates, as they indeed admit an hydrodynamic limit in which the acoustic perturbations can be described by an acoustic metric as for a perfect fluid \cite{garay2000becbh,garay2001becbh,blv2001becbh}. The interesting thing about BECs, besides the possibility of well-controlled experimental realization, is that away from the hydrodynamic limit they have a well-known microscopic behavior, hence providing a UV completion of the fluid description. This is important to address questions about the propagation of fields in curved spacetimes, in particular the influence of UV (trans-Planckian) frequencies on semiclassical effects such as the Hawking radiation. 

Also in the relativistic case we have a realization of the analogy in a quantum system, that is a relativistic Bose--Einstein condensate (RBEC). The acoustic metric describing perturbations around the background is of the same form as the one for a relativistic perfect fluid, and this is essentially given by the existence of an hydrodynamic limit for the relativistic condensate. The use of these systems as gravitational analogues was presented in \cite{rbec2010}; we give here a brief overview of the characteristics of RBECs we will need. 

The model is defined through a Lorentz invariant Lagrangian density for an interacting charged scalar field on a flat (Minkowskian) background
\begin{equation}\label{eq:rbec-lagrangian}
\hat{\mathcal{L}}=-\de_\mu\hat\p^*\de^\mu\hat\p-\left(\frac{m^2c^2}{\h^2}+V\right)\hat\p^*\hat\p-U(\hat\p^*\hat\p),
\end{equation}
where $m$ is the mass of the bosons, $V(t,\vc{x})$ is an external potential and $U$ is an interaction term. The possibility of condensation is given by the conservation of the charge of the condensate, that is the difference between particles and antiparticles. In the condensed phase, as in the non-relativistic case, we can describe the condensate at the mean-field level with a scalar wavefunction $\phi$ satisfying
\begin{equation}\label{eq:rbec-gp}
\de_\m\de^\m \p -\left(\frac{m^2c^2}{\hbar^2}+V\right)\p-U'(\p^*\p)\p=0,
\end{equation}
where the prime indicates the derivative with respect to $\r$, while we can treat fluctuations as perturbations to the background as
\begin{equation}
\hat\p=\p(1+\hat\psi).
\end{equation}

For our scopes it is convenient to adopt the Madelung (density-phase) representation
\begin{equation}
\p=\sqrt{\rho}\ e^{i\t}
\end{equation}
and to introduce the quantities 
\begin{equation}\label{eq:rbec-definitions}
u_\m\defeq \frac{\hbar}{m}\de_\m\theta\ ; \quad c_0^2=\frac{\hbar^2}{2m^2}\r U'';\ \quad T_\r\defeq -\frac{\hbar^2}{2m\r}\de^\m\r \de_\m.
\end{equation}
With these definitions the charge current conservation associated to the U(1) symmetry of the Lagrangian can be expressed as
\begin{equation}\label{eq:rbec-continuity} 
  \de_\m j^\m=0\;;\hspace{1cm}j^\m\defeq\r u^\m,
\end{equation}
while equation (\ref{eq:rbec-gp}) takes the form
\begin{equation}\label{eq:rbec-bernoulli-nonapprox}
-u_\m u^\m=c^2+\frac{\hbar^2}{m^2}\left[V+U'(\r)-\frac{\de_\m\de^\m\sqrt{\r}}{\sqrt{\r}}\right].
\end{equation}
Instead the equation for the perturbation $\hat\psi$ is 
\begin{equation}\label{eq:rbec-perturb-equation}
\left(\left[i\hbar u^\m\de_\m+T_\r\right]\frac{1}{c_0^2}\left[i\hbar u^\m\de_\m-T_\r\right] -\frac{\hbar^2}{\r}\de^\m\r\de_\m\right)\hat\psi=0.
\end{equation}

\subsubsection{The hydrodynamic limit: the acoustic metric in RBECs}
\label{sec:rbec-hydrodynamic-lim}
We are now going to show that the equations describing the RBEC background admit a limit in which they become the equations for a relativistic inviscid, barotropic and irrotational fluid. A similar discussion, with a different notation, can be found in \cite{boisseau2004vortex}.

Equation (\ref{eq:rbec-continuity}) is already of the shape of a continuity equation, but we need to identify what the RBEC charge density $\r$ is from the point of view of a fluid. We can start by writing $\r$ as
\begin{equation}\label{eq:rbec-densities}
\r=\frac{n}{w},
\end{equation}
where $n$ and $w$ will turn out to correspond to the number density and the specific enthalpy of the fluid. 
We can now write the 4-gradient of the phase in terms of a unit 4-vector $V^\m$, i.e. such that $V_\m V^\m=-1$, that will be identified with the normalized 4-velocity of the perfect fluid,
\begin{equation}
\de_\m \theta=wV_\m,
\end{equation}
that corresponds to the following expression for the 4-current
\begin{equation}\label{eq:rbec-current-fluid}
j_\m=n V_\m.
\end{equation}
Thus the conserved current has the correct expression in terms of the would-be number density and velocity.

We can compare (\ref{eq:rbec-current-fluid}) and (\ref{eq:rbec-continuity}) to obtain the fluid 4-velocity in terms of the quantity $u_\m$ we defined for the RBEC
 \begin{equation}\label{eq:rbec-4vel}
V_\m=\frac{m}{\hbar}\frac{u_\m}{w}.
 \end{equation}
From this we can see that the one-form associated to the 4-velocity has the form ${v=\a(x)d\theta}$, that is it is irrotational in the relativistic sense. Moreover we can obtain a unit timelike vector field from $u_\m$ by normalizing it, that is we can also express the fluid velocity without referring to thermodynamic quantities as
\begin{equation}\label{eq:rbec-current-normalization}
V_\m=\frac{u_\m}{\norm{u}}\ ;\hsp \norm{u}\defeq\sqrt{-u_\sigma u^\sigma}.
\end{equation}

Now notice that equation (\ref{eq:rbec-bernoulli-nonapprox}) is reminiscent of Bernoulli equation (\ref{eq:bernoulli-equation}). In particular if the charge density $\r$ of the condensate varies slowly compared to the variations of the phase $\theta$ one can neglect the term $\frac{\de_\m\de^\m\sqrt{\r}}{\sqrt{\r}}$ and obtain a Bernoulli equation with the thermodynamic quantities determined by the mass and interaction terms of the Lagrangian. This is the hydrodynamic limit, in which the condensate behaves as an inviscid, barotropic and irrotational relativistic fluid.

Consider now equation (\ref{eq:rbec-perturb-equation}) describing the perturbations. The dispersion relation associated to it is much more complex than the Bogoliubov one for non-relativistic condensates and has more interesting regimes that were fully analyzed in \cite{rbec2010}. Here we are interested in the low momentum regime, given by the following condition on the modes wavenumber $k$
\begin{equation}
|k|\ll \frac{mu^0}{\hbar}\left[1+\left(\frac{c_0}{u^0}\right)^2\right],
\end{equation}
where the perturbations have a phononic behavior. We also restrict to modes with period of oscillation and wavelength much greater than the typical scales over which the background quantities vary, this is called \textit{eikonal approximation}. In this regime we can neglect the terms with $T_\r$ in (\ref{eq:rbec-perturb-equation}) so that, also using $\de_\m (\r u^\m)=0$, the equation describing the evolution of perturbations $\hat\psi$ can be rewritten as a massless Klein-Gordon equation
in a curved background specified by the acoustic metric
\begin{equation}
\G_{\m\n}=\frac{\r}{\sqrt{1-u_\sigma u^\sigma/c_0^2}}\left[\eta_{\m\n}\left(1-\frac{u_\sigma u^\sigma}{c_0^2}\right)+\frac{u_\m u_\n}{c_0^2}\right],
\end{equation}
that can be brought in the Gordon form in terms of $V_\m$ as written in (\ref{eq:rbec-current-normalization}) and of the speed of sound that is defined because of how it appears in the metric as $c_s^2\defeq c^2c_0^2/(\norm{u}^2+c_0^2)$.
The metric in Gordon form is 
\begin{equation}\label{eq:rbec-gordon-metric}
\G_{\m\n}=\r\frac{c}{c_s}\left[\eta_{\m\n}+\left(1-\frac{c_s^2}{c^2}\right)V_\m V_\n\right].
\end{equation}
Also the conformal factor is the same (up to an irrelevant constant factor) as the one of the metric (\ref{eq:relativistic-acoustic-metric}) for a relativistic fluid because of the identification (\ref{eq:rbec-densities}) of the condensate wavefunction amplitude $\r$.

The non-relativistic limit is here obtained when the normalized flow velocity $v^i$ and the speed of sound $c_s$ are much smaller than the speed of light $c$ and also the self interaction between atoms is weak, condition that is expressed by $c_0\ll c$. From (\ref{eq:rbec-bernoulli-nonapprox}) (in the hydrodynamic limit) in the limit of small interaction we are left with $\norm{u}^2=- u_\m u^\m\simeq -c^2$ that also implies $u^0\simeq c$. Moreover the speed of sound reduces to $c_s\simeq c_0$. Notice that we also get
\begin{equation}
V^\m=\frac{u^\m}{\norm{u}}\simeq\frac{u^\m}{c},
\end{equation}
so that $u^i$ reduces to the fluid flow velocity, without issues of normalization.

With these considerations, analogously to the case of a relativistic perfect fluid, one can show that the Gordon metric (\ref{eq:rbec-gordon-metric}) reduces to the one for a non-relativistic condensate.

\subsubsection{Fluid vs RBEC for analogue gravity}
For what we said the gravitational analogue with Gordon metric in RBECs makes sense only in the hydrodynamic limit, in which the condensate behaves as a relativistic fluid. Hence there should be no difference in the simulation of acoustic black holes.

However, because of boson-antiboson annihilation, for the condensate the most natural \textit{density} with respect to which to write the continuity equation is not the number density $n$ as in the case of a proper fluid, but the \textit{charge density} $\r$ that is related to the other one via (\ref{eq:rbec-densities}). This is relevant because assuming \textit{constant density} can be a sensible and simplifying assumption when dealing with \textit{canonical} analogue spacetimes (that is spacetimes that are not solutions of general relativity, but present the relevant features that one wants to reprduce, e.g.~analogue horizons) and for a RBEC the density that is expected to be naturally adjustable is the charge density $\r$.

This means that dealing with constant density $\r$ in condensates corresponds to a fluid description with non-constant number density, differently from the non-relativistic case in which the density of the condensate corresponds to the number density in the fluid description.

One can hence naturally get different analogue spacetimes considering an actual fluid and a RBEC, even if every metric that admits Gordon form can be obtained with both systems. Also, the calculations turn out to be simpler for a relativistic condensate of constant $\r$ than for a fluid of constant $n$ since the continuity equation involves $u^\m$ that is an exact 4-gradient and hence we have stricter conditions and do not need to enforce the irrotationality condition explicitly.


\section{Relativistic vs non-relativistic acoustic metrics}
\label{sec:rel-vs-nonrel}
We already said that in the non-relativistic limit, in which both the flow velocity and the speed of sound are much smaller than the speed of light, the relativistic acoustic metric reduces to the non-relativistic one. Now we are interested in understanding if the different structure of the Gordon metric permits us to simulate spacetimes different from the ones achievable with the non-relativistic acoustic metric.

As was pointed out in \cite{rbec2010} the spacetimes one can simulate with a non-relativistic acoustic metric are limited by the fact that they must have, for some choice of coordinates, conformally flat spatial slices. Thus we cannot obtain non-relativistic acoustic analogues of many spacetimes, including the Kerr spacetime. Indeed, it was proven that a wide class of stationary spacetimes does not admit conformally flat spatial slices (see for example \cite{kroon2004nonexistence,visser-wein-kerr-equator}).

The Gordon metric instead does not in general have conformally flat spatial slices and hence admits in principle the simulation of different acoustic spacetimes. Actually we are now going to show that the Gordon metric can simulate \textit{more general} stationary spacetimes, in the sense that every stationary metric that can be put in the non-relativistic acoustic form can also be put in the Gordon form.

We will now look for a transformation that brings us from a relativistic acoustic metric (or metric in the Gordon form) to a non-relativistic one and to see when it is valid. We focus on stationary flow configurations, that is with the velocity components independent from time, and assume the speed of sound to be constant.

We consider now the 2+1 dimensional case since the 3+1 dimensional one is analogous; also we consider the acoustic metrics up to a conformal factor since it is not relevant for the point we are trying to make.
Let's take a generic metric in the Gordon form. The associated line element is
\begin{equation}
\begin{split}
ds&^2=-(1-\xi V_0^2)c^2dt^2+2\xi V_0V_1 cdtdx_1+2\xi V_0V_2cdtdx_2\\
&+2\xi V_1V_2dx_1dx_2+(\eta_{11}+\xi V_1^2)dx_1^2+(\eta_{22}+\xi V_2^2)dx_2^2,
\end{split}
\end{equation}
where we indicate with subscripts 1 and 2 two generic spatial coordinates. The idea is to find a coordinate transformation that eliminates the spatial cross terms from the metric, leaving conformally flat spatial slices. This is achieved with the change of coordinates 
\begin{equation}\label{eq:rel-nonrel-2+1}
  dt'=dt\pm\frac{\sqrt{\xi}V_1}{1-\sqrt{\xi} V_0}dx_1 \pm \frac{\sqrt{\xi}V_2}{1-\sqrt{\xi} V_0}dx_2,
\end{equation}
with which the metric takes the desired form
\begin{equation}\label{eq:acoustic-metric-transformation}
\begin{split}
  ds^2=&-(1-\xi V_0^2)c^2dt'^2 \pm 2\sqrt{\xi}V_1 cdt'dx_1 \\
  &\pm 2\sqrt{\xi}V_2 cdt'dx_2 + \eta_{11}dx_1^2+\eta_{22}dx_2^2.
  \end{split}
\end{equation}
The $dt'^2$ term can be rewritten as
\begin{equation}
-(1-\xi V_0^2)c^2=-(1-\xi(1+\vc{V}^2))c^2=-(c_s^2-c^2\xi\vc{V}^2)\, ,
\end{equation}
where $\vc{V}$ is here the one-form associated to the spatial part of the velocity.

Now remember that the covariant 4-vector $V_\mu$ is the fluid covariant 4-velocity divided by $c$, hence we can define the covariant velocity components
\begin{equation}\label{eq:corresp-nonrel-velocity-def}
w_i\defeq c\sqrt{\xi}V_i,
\end{equation}
that are related to the velocity vector components $\tld{w}^i$ with respect to a normalized tangent space basis by
\begin{equation}\label{eq:nonrel-rel-transf-vel}
\tld{w}^i=\frac{w_i}{h_i}=\frac{c\sqrt{\xi}}{h_i}V_i,
\end{equation}
where $h_i\defeq \sqrt{\eta_{ii}}$ are the scale factors.

In terms of this velocity $w$ the form (\ref{eq:acoustic-metric-transformation}) of the metric has the manifest form of a non-relativistic acoustic metric
\begin{eqnarray}
ds^2&=&-(c_s^2-\vc{\tld{w}})^2dt'^2 \pm 2 h_1 \tld{w}^1dt'dx_1 \pm 2h_2 \tld{w}^2dt'dx_2 \nonumber\\
&&+ \eta_{11}dx_1^2+\eta_{22}dx_2^2.
\end{eqnarray}

The 3+1 dimensional case is analogous and the change of coordinates (\ref{eq:rel-nonrel-2+1}) takes the general form
\begin{equation}\label{eq:rel-nonrel-transf}
dt'=dt\pm\frac{\sqrt{\xi}V_i}{1-\sqrt{\xi} V_0}dx_i
\end{equation}
with a sum over the spatial indeces $i$.

The question is now for which flows $V_\mu$ this change of coordinates is integrable. We can find a condition imposing for the one-form that defines the change of coordinates to be closed, i.e. if the change of coordinates is $dt'=\omega_\mu dx^\mu$ (where $dx^\mu$ are a basis of one-forms and $dt'$ could also not be an exact differential) we have to impose
\begin{equation}
d\omega=0.
\end{equation}
For both the (2+1) and the (3+1)-dimensional case, using also the relativistic irrotationality condition and remembering the assumptions of constant speed of sound $c_s$ and stationary flow, this condition becomes 
\begin{equation}\label{eq:ac-metrics-transf-int-condition}
  \d \vc{V}=\de_{[i}V_{j]}=0.
\end{equation}
Hence if the velocity flow satisfies this condition the change of coordinates is well defined globally and we can bring the relativistic acoustic metric into a non-relativistic one.

This integrability condition implies that the transformation is not possible for every relativistic acoustic metric derived for an irrotational flow, since (\ref{eq:ac-metrics-transf-int-condition}) is not in general satisfied when the irrotationality condition $V\wedge dV$ is. Hence, as expected, we cannot bring any acoustic Gordon metric to a non-relativistic acoustic one, and this is an indication of the fact that the relativistic acoustic analogues admit the simulation of different spacetimes. Moreover notice that, when the transformation is applicable, the non-relativistic flow (\ref{eq:corresp-nonrel-velocity-def}) corresponding to the 4-velocity of a relativistic fluid is not in general its non-relativistic limit.

However, if we consider the transformation in the other direction
\begin{equation}\label{eq:nonre-rel-transf-inv}
dt'=dt\pm \frac{h_i \tld{w}_i}{c-\sqrt{c^2-c_s^2+\vc{v}^2}}dx_i,
\end{equation}
the integrability condition (\ref{eq:ac-metrics-transf-int-condition}), also because $\xi$ is assumed constant,  corresponds to the vanishing of the rotor of the velocity (\ref{eq:corresp-nonrel-velocity-def}). Since the non-relativistic acoustic metric is derived for an irrotational fluid we can hence say that any stationary non-relativistic acoustic metric can be put in Gordon form and hence in principle (from the geometric point of view) \textit{all the stationary spacetimes that can be simulated in non-relativistic fluid analogues (with constant speed of sound) can also be obtained in a relativistic fluid}. In this sense the Gordon form is a more general form for a stationary metric than the non-relativistic acoustic one.

\subsection{An example: the Schwarzschild black hole}
\label{sec:schwarzschild-gordon-form}
We can use the transformation we just found to obtain the Schwarzschild spacetime in Gordon form up to a conformal factor, and indeed the non-relativistic acoustic simulation of its causal structure is known and was presented in \cite{visser-acoustic-bh}. In particular the Schwarzschild line element written in Painlevé--Gullstrand coordinates
\begin{equation}
  ds^2=-dt_{PG}^2+\left(dr\pm \sqrt{\frac{2GM}{r}}dt_{PG}\right)^2+r^2(d\t^2+\sin^2\t d\p^2)
\end{equation}
has the shape of a non-relativistic acoustic metric with a radial velocity 
\begin{equation}
\tld{w}_r=\sqrt{\frac{2GM}{r}}.
\end{equation}

We now need to apply to the Painlevé--Gullstrand line element the inverse transformation with $t=t_{PG}$, that in this case takes the form
\begin{equation}
dt'=dt_{PG}\pm \sqrt{\frac{2GM}{r}}\frac{1}{c-\sqrt{c^2-c_s^2+\vc{w}^2}}dr.
\end{equation}

As we showed for the general case above this gives us a Gordon metric with a radial component of the unit 4-velocity given by (\ref{eq:nonrel-rel-transf-vel}) and $V_0$ given by normalization, so that in spherical coordinates
\begin{equation}\label{eq:schwarzschild-rel-analogue-vel}
V_\m=\left(\sqrt{1+\frac{2GM}{\xi c^2r}}, \sqrt{\frac{2GM}{\xi c^2r}},0,0\right).
\end{equation}
Hence the Gordon metric with this $V_\m$ is, up to a conformal factor, the Schwarz\-schild one in unusual coordinates with $c$ substituted by $c_s$. 

For a relativistic fluid we need to impose the continuity equation in spherical coordinates, that is
\begin{equation}
\de_r\left(r^2 n V^r\right)=0\ \Longrightarrow\ nV^r\propto \frac{1}{r^2},
\end{equation}
that implies for the number density
\begin{equation}
n\propto \sqrt{\frac{c^2\xi}{2GM}}.
\end{equation}

We can compare this result with the RBEC configuration that gives rise to a Gordon metric conformally related to Schwarzschild that was presented in \cite{rbec2016}.
By normalizing the 4-current field obtained there one obtains the 4-velocity field (\ref{eq:schwarzschild-rel-analogue-vel}) we found.

\vspace{0.5cm}

\section{Rotating acoustic black holes}
\label{sec:rotatingacoustic}
In \cite{rbec2016} it was shown how static metric solutions can be incorporated in the formalism of relativistic acoustic spacetimes. We now focus on stationary axisymmetric solutions, having in mind rotating black holes.

We start by considering a transformation of the Gordon metric that brings it into a form that is manifestly stationary and axisymmetric. What we want to obtain is something of the form 
\begin{equation}\label{eq:general-axisymmetric}
ds^2=g_{tt}dt^2+2g_{t\p}dt\,d\p+g_{11}dx_1^2+g_{22}dx_2^2+g_{\p\p}d\p^2,
\end{equation}
that is a fairly general stationary, axisymmetric and asymptotically flat metric (see for example \cite{wald-gr,hobson}), for example the Kerr rotating solution of the Einstein equations written in Boyer--Lindquist coordinates has this form. Here $t$ and $\p$ are Killing parameters, that is $(\de/\de t)^\m$ and $(\de/\de\p)^\m$ are Killing vectors, and $x_1,x_2$ are appropriate coordinates in the planes orthogonal to the ones spanned by the Killing vectors.

Take the Gordon metric in flat background, neglecting the conformal factor, written in generic curvilinear coordinates $(t,x_1,x_2,\p)$:
\begin{equation}\label{eq:gordon-metric-generic-coordinates}
\begin{split}
ds^2=&-(1-\xi V_0^2)c^2dt^2+2\xi V_0V_1 cdt\, dx_1+2\xi V_0V_2 cdt\, dx_2\\
& + 2\xi V_0V_\p cdt\, d\p+(1+\xi V_1^2)dx_1^2\\
&+ 2\xi V_1V_2 dx_1\, dx_2 + 2\xi V_1V_\p dx_1\, d\p + (\eta_{22}+\xi V_2^2)dx_2^2\\
&+2\xi V_2V_\p dx_2\, d\p+(\eta_{\p\p}+\xi V_\p^2)d\p^2,
\end{split}
\end{equation}
where $\eta_{ii}$ are the elements of Minkowski metric. The coordinates $x_1,x_2$ can for example be $r,\t$ in spherical coordinates, or $r,z$ in cylindrical coordinates. We take the unit timelike field $V_\m$ to be independent from $t$ and $\p$ since one expects an axisymmetric metric to have components independent from the coordinates that are Killing parameters.

The change of coordinates we are looking for is given by (where we indicate the Minkowski metric component $\eta_{\p\p}$ with $\eta_\p$)
\begin{equation}\label{eq:gordon-axisymm-coor-change}
  \begin{dcases}
	\!\begin{aligned}
	  cdt'=cdt&-\frac{\eta_\p\xi V_0V_1}{\xi V_\p^2+\eta_\p(1-\xi V_0^2)}dx_1\\
	  &\hsp- \frac{\eta_\p\xi V_0V_2}{\xi V_\p^2+\eta_\p(1-\xi V_0^2)}dx_2
  	\end{aligned}
	\\
	\!\begin{aligned}
	  d\p'=d\p&+\frac{\xi V_1V_\p}{\xi V_\p^2+\eta_\p(1-\xi V_0^2)}dx_1\\
	  &\hsp+\frac{\xi V_2V_\p}{\xi V_\p^2+\eta_\p(1-\xi V_0^2)}dx_2
  	\end{aligned}
	\\
	dx_1'=dx_1+\frac{\xi\eta_{22} V_1V_2}{\eta_{22}(1-\xi)-\xi V_2^2}dx_2.
  \end{dcases}
\end{equation}

The condition of integrability is not very compact and hence it is more convenient to check it with specific $V$-s. The Gordon metric with respect to coordinates $(t',x_1',x_2,\p')$ assumes the desired form
\begin{widetext}
\begin{equation}\label{eq:rel-acoustic-bl}
\G_{\mu\nu}\propto
\begin{bmatrix}
 -(1-\xi V_0^2)c^2	& 0	& 0	& \xi V_0V_\p \\
 0 & {\displaystyle 1+\frac{\xi\eta_\p V_1^2}{\xi V_\p^2+\eta_\p(1-\xi V_0^2)}}	& 0	& 0 \\
 0 & 0	& {\displaystyle \eta_{22}+\frac{\xi\eta_{22}V_2^2}{\eta_{22}(1-\xi)-\xi V_2^2}}	& 0 \\
 \xi V_0 V_\p 		& 0	& 0	& \eta_\p+\xi V_\p^2 
\end{bmatrix}.
\end{equation}
\end{widetext}

This form of the metric shows that stationary acoustic spacetimes can be obtained in relativistic gravitational analogues, and in particular rotating spacetimes, with the information of the rotation encoded in the azimuthal velocity $V_\p$, in fact when this vanishes the metric assumes a manifestly static (diagonal) form.

To show that rotating acoustic black holes are possible we must show that horizons for this kind of metric can occur. Remember that the event horizon is a null hypersurface and, for the symmetries of the spacetime, it will be determined by the vanishing of some scalar quantity depending only on $x_1$ and $x_2$, that is
\begin{equation}
f(x_1,x_2)=0 \hspace{.5cm}\mbox{with}\hspace{.5cm} \G^{\m\n}\de_\m f\de_\n f=0,
\end{equation}
that for the axisymmetric metric we are considering becomes
\begin{equation}
\G^{11}(\de_1 f)^2 + \G^{22}(\de_2 f)^2=0.
\end{equation}
We are however free to redefine the coordinates on the 2-planes orthogonal to the Killing fields $(x_1,x_2)\to(\tld{x}_1,\tld{x}_2)$ so that the above condition reduces to
\begin{equation}\label{eq:axisymm-horizon-conditon}
\G^{11}(\de_1 f)^2=0,
\end{equation}
that corresponds to saying that the event horizon occurs when $\G^{11}=0$, or when $\G_{11}$ diverges. From (\ref{eq:rel-acoustic-bl}) this condition reads (remember $V_\m=v_\m/c$)
\begin{equation}\label{eq:relativisti-stationary-horizon}
\frac{v_1^2}{\eta_{11}}+\frac{v_2^2}{\eta_{22}}=\frac{c_s^2}{\xi}.
\end{equation}

Another interesting location for a rotating black hole is the ergosurface that is found from the vanishing of the $\G_{00}$ component. In particular from (\ref{eq:rel-acoustic-bl}) the condition that individuates the ergosurface is 
\begin{equation}\label{eq:relativisti-stationary-ergo}
\vc{v}^2=\frac{v_1^2}{\eta_{11}}+\frac{v_2^2}{\eta_{22}}+\frac{v_\p^2}{\eta_\p}=c^2\frac{1-\xi}{\xi}=\frac{c_s^2}{\xi}.
\end{equation}

Remember that, as explained in \cite{visser-acoustic-bh}, in non-relativistic analogue gravity the ergosurface coincides with the surface where the fluid flow turns supersonic, while the horizon is a surface on which the normal component of the flow becomes supersonic. Conditions (\ref{eq:relativisti-stationary-horizon}) and (\ref{eq:relativisti-stationary-ergo}) are the relativistic generalizations of these notions, in fact they differ from the non-relativistic ones by a relativistic Lorentz factor $\xi^{-1}=(1-c_s^2/c^2)^{-1}=\gamma(c_s)^2$. Notice that, coherently with what we said above about the role of $V_\p$, the ergosurface is distinct from the horizon only if $v_\p\neq 0$.

\subsection{(2+1)-dimensional stationary acoustic spacetimes}
As one can see from (\ref{eq:relativistic-acoustic-metric}) the Gordon metric depends on the dimensionality of the spacetime only in the conformal factor. Hence if we are interested in the causal structure of (2+1)-dimensional spacetimes with polar symmetry we can use cylindrical coordinates in the argument of the previous section, that is take $(x_1,x_2)=(r,z)$, impose $V_2=V_z=0$ and work on constant-$z$ planes.
One then obtains the relativistic acoustic metric, with respect to suitable coordinates $(t',r,\p')$, in the form
\begin{equation}\label{eq:metric-cyl}
\G_{\mu\nu}\propto
\begin{bmatrix}
 -\left(1-\xi V_0^2\right)c^2	& 0	& \xi V_0V_\phi c\\
 0		& 1+\frac{\xi r^2 V_r^2}{\xi V_\phi^2+r^2(1-\xi V_0^2)}	& 0\\
 \xi V_0V_\phi c	& 0	& r^2+\xi V_\phi^2
\end{bmatrix}.
\end{equation}

For the position of the horizon and the ergosurface (that will now be circles) the same discussion we did for the (3+1)-dimensional case holds. Moreover, notice that the polar symmetry and the stationarity require for the velocity to depend only on the radial coordinate $x_1=r$.


\section{Relativistic vortex geometry}
\label{sec:vortex}
In non-relativistic analogue gravity the simplest example of rotating spacetime is the vortex geometry presented in \cite{visser-acoustic-bh} obtained for a constant density fluid moving with flow speed
\begin{equation}\label{eq:nonrel-vortex-flow}
  \vc{v}=\frac{A}{r}\hat{\vc{e}}_r+\frac{B}{r}\hat{\vc{e}}_\phi,
\end{equation}
that has an associated acoustic metric
\begin{equation}
\begin{split}
  ds^2=&-\left(c_s^2-\frac{A^2+B^2}{r^2}\right)dt^2\\
  &-2\frac{A}{r}dt\ dr-2B dt\ d\phi+dr^2+r^2d\phi^2.
  \end{split}
\end{equation}

\subsection{Vortex in a constant density relativistic fluid}
We are now interested in the relativistic case. Consider a constant density relativistic perfect fluid; we are looking for (2+1)-dimensional configuration with circular symmetry satisfying the relativistic continuity equation~(\ref{eq:rel-continuity}) and the relativistic irrotationality condition (\ref{eq:rel-irrotationality}). For what concerns the Bernoulli equation (\ref{eq:bernoulli-equation}) we can use the freedom on the equation of state to satisfy it with the flow we desire.

Because of the requests of stationarity and circular symmetry the velocity components can only depend on the radial coordinate, so that the continuity equation is
\begin{equation}
  \frac{1}{\sqrt{|\eta|}}\de_\mu(\sqrt{|\eta|} v^\mu)=\frac{1}{r}\de_r(rv^r)=0 \implies v^r=v_r=c_s\frac{r_0}{r},
\end{equation}
where $r_0$ is a constant, while the irrotationality condition is
\begin{equation}\label{eq:irrot-cond-fluid-vortex}
  v_0\de_r v_\phi - v_\phi \de_r v_0=0.
\end{equation}
We can now insert the normalization condition for the 4-velocity
\begin{equation}
  v_0=\pm \sqrt{c^2+v_r^2+\frac{v_\phi^2}{r^2}}=\sqrt{c^2+\frac{c_s^2r_0^2+v_\phi^2}{r^2}}
\end{equation}
and integrate equation (\ref{eq:irrot-cond-fluid-vortex}) to obtain $v_\p$. We would like for the resulting Gordon metric to be asymptotically flat. We already saw that the radial velocity goes to zero at infinity, so we require also for the azimuthal velocity to vanish asymptotically. Note however, that $v_\phi(\infty)=0$ is not the correct condition since it is a component of the covariant 4-velocity and what we want to set to zero is the component $\tld{v}_\p=v_\p/r$ with respect to an orthonormal basis of the tangent vector space. Hence we need to impose for $v_\p$ to be a constant at infinity, define
\begin{equation}
ac_s\defeq v_\phi(\infty)=\mathrm{const}.
\end{equation}
With this initial condition the integration of the irrotationality condition gives
\begin{equation}
  v_\phi^2=K^2\frac{c^2r^2+c_s^2r_0^2}{c^2r^2-c_s^2a^2}.
\end{equation}
Hence the covariant 4-velocity of the fluid is
\begin{equation}\label{eq:rel-vortex-fluid-flow}
v_\mu=\left(c\sqrt{\frac{c^2r^2+c_s^2r_0^2}{c^2r^2-c_s^2a^2}}\ ,\ c_s\frac{r_0}{r}\ ,\ c_sa\sqrt{\frac{c^2r^2+c_s^2r_0^2}{c^2r^2-c_s^2a^2}}\right).
\end{equation}

\begin{figure}[ht]
  \centering
  \includegraphics[width=0.3\textwidth]{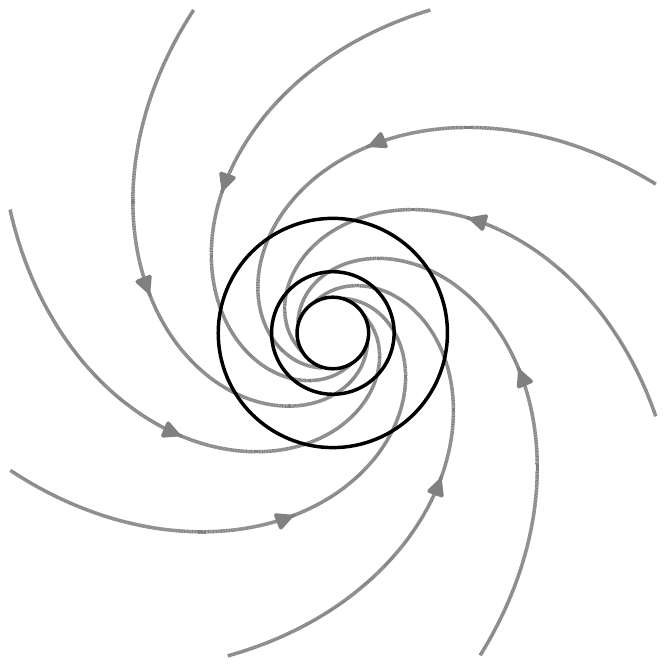}
 \caption{\small\textit{Representation of the streamlines of the spatial part of the relativistic vortex velocity field (\ref{eq:rel-vortex-fluid-flow}) for $a=1.5r_0$ and $\xi=0.9$. For this choice of the parameters there is an horizon. The innermost circle is the singularity radius $r_S$ and in the region it delimits the flow is not well defined. The other two circles are, in order, the horizon $r_H$ and the ergosurface $r_E$.}}\label{fig:rel-vortex-streamlines}
\end{figure}

Inserting this in the metric (\ref{eq:metric-cyl}) we obtain the line element
\begin{widetext}
\begin{equation}\label{eq:vortex-rel-fluid-bl}
ds^2=-\left(\frac{r^2c^2-\xi r_0^2 c^2-a^2c^2}{r^2c^2-a^2c_s^2}\right)c_s^2dt'^2+
2\xi a\frac{r^2c^2+c_s^2r_0^2}{r^2c^2-a^2c_s^2}\ c_s dt'd\phi'+\frac{r^2}{r^2-\xi r_0^2}dr^2 + \left(r^2+a^2\frac{c_s^2}{c^2}\frac{r^2c^2+r_0^2c_s^2}{r^2c^2-a^2c_s^2}\right) d\phi'^{\;2}.
\end{equation}
\end{widetext}

We can find the positions of the ergosurface $r_E$ and the horizon $r_H$ by using conditions (\ref{eq:relativisti-stationary-horizon}) and (\ref{eq:relativisti-stationary-ergo}) or by checking where the $dt'^2$ term of the metric vanishes and where the $dr^2$ term diverges. The results are
\begin{equation}
  r_E^2\defeq\xi r_0^2+a^2\;; \hspace{1cm} r_H^2\defeq\xi r_0^2.
\end{equation}
Notice also that the metric is singular for
\begin{equation}
r_S^2\defeq a^2\frac{c_s^2}{c^2},
\end{equation}
where the flow (\ref{eq:rel-vortex-fluid-flow}) becomes ill-defined, and hence the flow configuration has an acoustic horizon if $r_S<r_H$ that is if $a^2<\frac{\xi}{1-\xi}r_0^2$. This condition is analogous to the condition $a<M$ for the Kerr geometry to avoid naked singularities, however here we do not have the double horizon structure as in Kerr.

Notice that the fact that the singularity is a ring differs from the non-relativistic case, in which the flow is well defined till $r=0$. Also, when the azimuthal velocity goes to zero, i.e. $a\to0$, the horizon and the ergosurface coincide as for a static black hole and the singularity of the flow moves to $r_S=0$. We also observe that even if the term $dt'd\p'$ in (\ref{eq:vortex-rel-fluid-bl}) is asymptotically constant, this effective spacetime is asymptotically flat and it can be seen by normalizing the angular coordinate. This was expected since we required this in fixing the initial condition for the azimuthal velocity.

\subsubsection{The non-relativistic limit}
Here the non-relativistic limit is given by the conditions
\begin{equation}
  c_s,\frac{r_0c_s}{r},\frac{ac_s}{r}\ll c
\end{equation}
that for the square root in the 4-velocity (\ref{eq:rel-vortex-fluid-flow}) implies
\begin{equation}
  \sqrt{\frac{c^2r^2+A^2}{c^2r^2-K^2}}= 1+\O\left(\frac{v_r^2}{c^2},\frac{a^2c_s^2}{r^2c^2}\right),
\end{equation}
so that
\begin{equation}\label{eq:vortex-fluid-nonrel-lim}
  v^\mu\simeq\left(c,\ \frac{r_0c_s}{r}, \frac{ac_s}{r^2}\right)
\end{equation}
This is to be compared with the velocity field (\ref{eq:nonrel-vortex-flow}) of the non-relativistic vortex geo\-me\-try, but we need to pay attention to the usual issue of the choice of the vector basis. The spatial part of (\ref{eq:vortex-fluid-nonrel-lim}) with respect to an orthonormal basis $(\hat{\vc{e}}_r,\hat{\vc{e}}_\phi)$ is
\begin{equation}
\vc{v}=\frac{r_0c_s}{r}\hat{\vc{e}}_r+\frac{ac_s}{r}\hat{\vc{e}}_\phi,
\end{equation}
that is exactly as (\ref{eq:nonrel-vortex-flow}).

\subsection{Vortex in a constant density RBEC}
Consider now a relativistic Bose--Einstein condensate with constant charge density $\r$. We need to enforce the continuity equation (\ref{eq:rbec-continuity}) and the irrotationality condition is substituted by the fact that the 4-current $u^\m$ must be a 4-gradient. Again we can use the freedom on the choice of the interactions of the Lagrangian to take the speed of sound to be constant and to satisfy the condition (\ref{eq:rbec-bernoulli-nonapprox}) on the norm of $u_\m$ for the flow we need.

Again, because of the interest in spacetimes with polar symmetry, the radial and azimuthal components of $u^\m$ must dependent only on $r$. The continuity equation for constant density, as for the fluid case, hence gives
\begin{equation}
\frac{1}{r}\de_r(ru_r)=0 \Longrightarrow u_r=\frac{r_0c_s}{r},
\end{equation}
while the fact that $u_\mu$ is a 4-gradient implies that $u_0$ and $u_\p$ cannot have space-temporal dependence, otherwise the radial component would depend on $t,\p$. So we take
\begin{equation}
u_0=\mathrm{const} \hsp u_\p=:ac_s=\mathrm{const}.
\end{equation}
Hence we have the 4-current
\begin{equation}\label{eq:vortex-rbec-current}
u_\mu=\left(u_0\ ,\ \frac{c_sr_0}{r}\ ,\ a c_s\right),
\end{equation}
so that the normalized covariant 4-velocity that enters the Gordon metric is
\begin{equation}
  V_\mu=\frac{1}{\sqrt{r^2-c_s^2(r_0^2+a^2)/u_0^2}} \left(r ,\ \frac{r_0c_s}{u_0}\ ,\ \frac{ac_sr}{u_0}\right) 
\end{equation}

The acoustic line element in the form (\ref{eq:metric-cyl}) with this flow becomes
\begin{widetext}
\begin{equation}\label{eq:vortex-rbec-bl}
  \begin{split}
	ds^2=-\left(\frac{r^2u_0^2-r_0^2c^2-a^2c^2}{r^2u_0^2-r_0^2c_s^2-a^2c_s^2}\right)c_s^2dt'^2
+&\frac{a\xi r^2c^2 c_s }{r^2u_0^2-r_0^2c_s^2-a^2c_s^2}dt' d\p'+\\
&\hspace{1cm}+\left(\frac{r^2u_0^2 -r_0^2c_s^2-a^2c_s^2}{r^2u_0^2 -r_0^2c^2-a^2c_s^2}\right) dr^2
+\left(1+\frac{\xi a^2c_s^2}{r^2u_0^2-r_0^2c_s^2-a^2c_s^2}\right)r^2 d\p'^2,
  \end{split}
\end{equation}
\end{widetext}
from which we can as before see that there is an ergosurface at
\begin{equation}
r_E\defeq\frac{c}{u_0}\sqrt{r_0^2+a^2},
\end{equation}
and an horizon at
\begin{equation}
r_H\defeq\frac{c}{u_0}\sqrt{r_0^2+a^2\frac{c_s^2}{c^2}},
\end{equation}
that, as before, could also have been obtained by studying the modulus of the 4-flow $V_\m$.

Also in this case, the ergosurface is always external to the horizon (since $c_s/c<1$), and coincides with it in the limit of vanishing azimuthal velocity, i.e.~for $a\to 0$. 
The metric is singular where the flow is so, that is for 
\begin{equation}
r_S\defeq\frac{c_s}{u_0}\sqrt{r_0^2+a^2},
\end{equation}
which is always smaller than $r_H$ for any value of $r_0$ and $a$. It is perhaps worth noticing that the radius of the singularity does not shrink to zero when the rotation vanishes as was the case for the relativistic fluid. Of course, a relativistic fluid analogue could reproduce the same feature but at the cost of allowing a non-constant number density.

Finally, let us stress that even if this vortex configuration is different from the one we derived above for a constant (number) density relativistic fluid, they both reduce to the vortex geometry (\ref{eq:nonrel-vortex-flow}) in the non-relativistic limit. 


\section{(3+1)-dimensional stationary acoustic black holes}
\label{sec:3vortex}
We showed in the general discussion of Section \ref{sec:rotatingacoustic} that the Gordon metric admits stationary acoustic black holes also in (3+1) dimensions. We are now going to show that it is reasonably easy to find an explicit 3-dimensional flow configuration for the scope; however, both the flow and the metric have generally complicated forms and are not as easy to understand as the vortex geometries we presented in the previous section. For simplicity, we shall focus on constant-density relativistic condensates.

\subsection{Zero polar flow}
Let us consider first a configuration for a RBEC with zero polar flow, that is with $u_\theta=u_2=0$. This corresponds to $V_2=0$. The continuity equation and the fact that $u_\m$ must be a 4-gradient imply for the 4-current
\begin{equation}
  u_\mu=\left(u_0\ ,\ -c_s \frac{r_0^2}{r^2}\ ,\ 0\ ,\ c_s a\right),
\end{equation}
where $r_0$ and $a$ are constants. This corresponds to a unit fluid 4-velocity $V_\mu=u_\mu/\norm{u}=v_\m/c$
\begin{equation}\label{eq:rbec-zero-polar-flow}
  V_\mu=\frac{1}{\sqrt{r^4-\frac{c_s^2 r_0^4}{u_0^2}-\frac{a^2 c_s^2 r^2}{u_0^2\sin^2\t}}} \left(r^2\ ,\ -\frac{r_0^2 c_s}{u_0}\ ,\ 0\ ,\ \frac{a c_s r^2}{u_0}\right).
\end{equation}
The metric is quite long to write down and the more compact form (\ref{eq:rel-acoustic-bl}) is not applicable here since the change of coordinates (\ref{eq:gordon-axisymm-coor-change}) is not integrable. Hence in order to determine the positions of the horizon and the ergosurface we have to check where respectively the radial part and the norm of the spatial part of the 4-current exceed $c_s/\xi$.

The horizon turns out to be located at the radius
\begin{equation}
r_H^2=\frac{c_s^2a^2}{2u_0^2\sin^2\t}+ \sqrt{\left(\frac{c_s^2a^2}{2u_0^2\sin^2\t}\right)^2+c^2\frac{r_0^4}{u_0^2}},
\end{equation}
while the ergosurface at
\begin{equation}
r_E^2= \frac{c^2a^2}{2u_0^2\sin^2\t}+ \sqrt{\left(\frac{c^2a^2}{2u_0^2\sin^2\t}\right)^2+c^2\frac{r_0^4}{u_0^2}}.
\end{equation}
Moreover, the metric is singular where the normalized 4-velocity (\ref{eq:rbec-zero-polar-flow}) is singular, that is at
\begin{equation}
r_S^2=\frac{c_s^2a^2}{2u_0^2\sin^2\t}+ \sqrt{\left(\frac{c_s^2a^2}{2u_0^2\sin^2\t}\right)^2+c_s^2\frac{r_0^4}{u_0^2}}.
\end{equation}

Notice that, due to the appearances of $c$ and $c_s$, we correctly have, for every value of $r_0$ and $a$,
\begin{equation}
r_S^2<r_H^2<r_E^2
\end{equation}
and that in the limit $a\to0$ both the horizon and the ergosurface move to the position of the horizon of the canonical static acoustic black hole presented in \cite{rbec2016}.

\subsection{Complete flow}
\begin{figure}[h]
\centering
\includegraphics[width=0.48\textwidth]{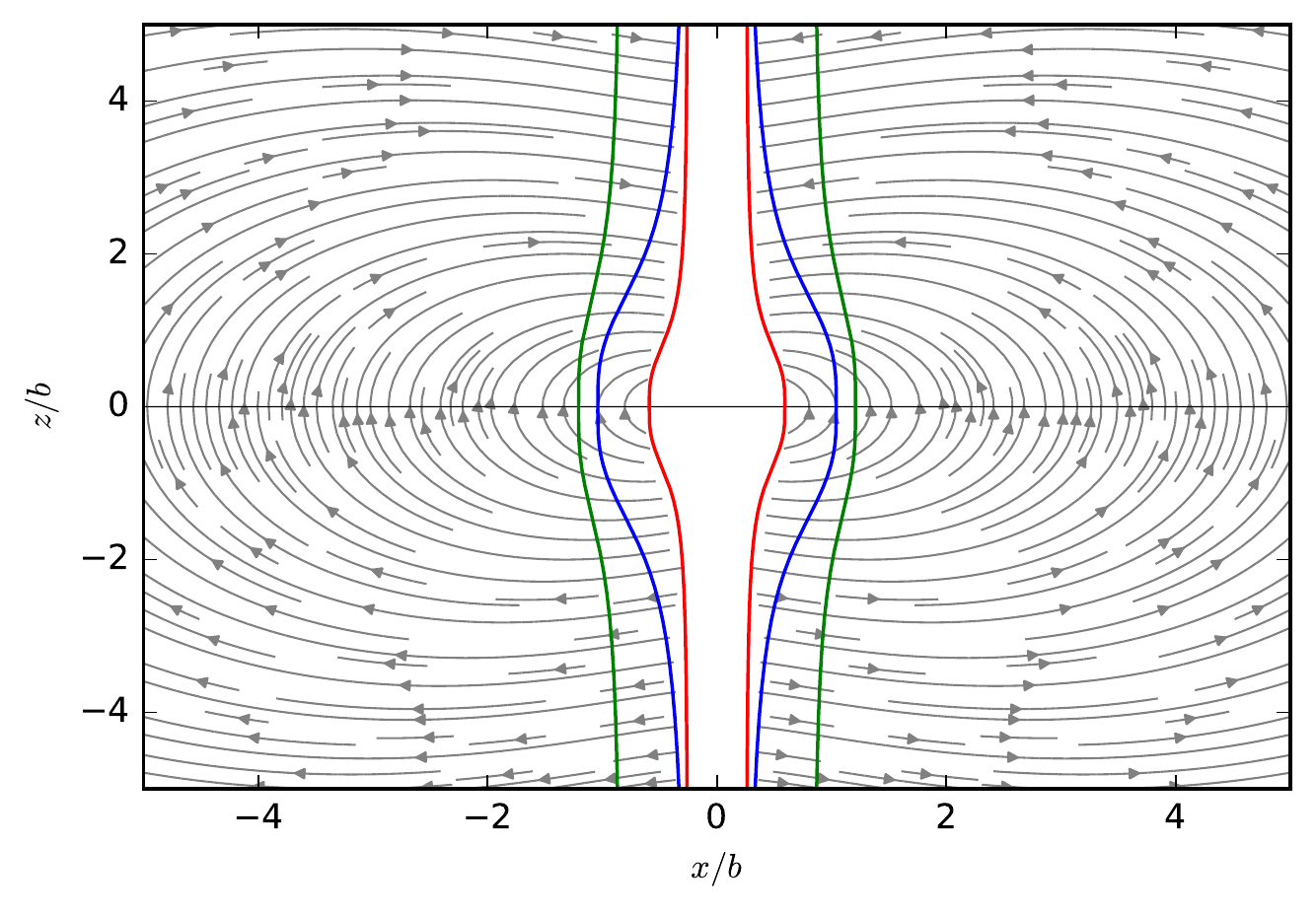}
\caption{\small\textit{Representation of the streamlines of the normalized 4-current for $\xi=0.9$, $u_0=c$, $r_0=0$ and $a=0.8b$ (expressed with respect to an orthonormal vector basis) on a vertical slice of the space (that is at $\p=const$). In the empty region at the center, delimited by the red line, the flow is ill-defined. The green line is the ergosurface and the blue one is the acoustichorizon. One can see that above and below the equator the streamlines pass through the horizon in opposite directions, meaning that in the southern hemisphere we have a \textit{white hole} horizon and in the northern hemisphere a black hole one.}}\label{fig:3+1-vortex-full-2}
\end{figure}

As above we are interested in stationary axisymmetric acoustic black holes, hence we look for a 4-current field that, in spherical coordinates, depends only on the radial and the polar coordinates. The constant-density continuity equation is
\begin{equation}
\de_r(r^2v_r)+\frac{1}{\sin\t}\de_\t(\sin\t\ v_\t)=0.
\end{equation}
A simple case is the one in which this is satisfied by the separate vanishing of the two terms, that is
\begin{equation}
u_r=\frac{A}{r^2}+\frac{B(\t)}{r^2}\ ; \hsp u_\t=\frac{D(r)}{\sin\t}.
\end{equation}
The other two components of the 4-current are again constant because of the 4-gradient constraint:
\begin{equation}
u_0=\mathrm{const} \hsp u_\p=\mathrm{const}=:c_s a.
\end{equation}
The fact that $u_\mu$ is a 4-gradient also implies that its differential must vanish, condition that, because of the sole $r$ and $\t$ dependence, only consists in the equality
\begin{equation}
\de_r u_\t-\de_\t u_r=0\ \Longrightarrow\ \frac{\de_r D(r)}{\sin\t}-\frac{\de_\t B(\t)}{r^2}=0,
\end{equation}
that can be integrated to obtain (up to additive constants)
\begin{equation}
D(r)\propto \ln\left(\frac{\cos\t+1}{\sin\t}\right)\ ;\hspace{1cm} B(\t)\propto 1/r,
\end{equation}
and also implies that these two functions are proportional to the same parameter, that we express as $c_s b^2$. Hence, introducing also $r_0^2\defeq A/c_s$, we have the 4-current
\begin{eqnarray}\label{eq:3d-flow-current}
u_\m&=&\left(u_0,u_r,u_\t,u_\p\right) \nonumber\\
&=& c_s\left(\frac{u_0}{c_s}, \frac{r_0^2}{r^2}+\frac{b^2}{r^2}\ln\left(\frac{\cos\t+1}{\sin\t}\right),\frac{b^2}{r\sin\t}, a\right).
\end{eqnarray}

Once normalized this 4-current can be put in the relativistic acoustic metric and the positions of the horizon and the ergosurface can be found as above. We do not write this down explicitly since the expressions are rather messy and not very significant. We instead give a graphical representation in Figure \ref{fig:3+1-vortex-full-2} of the streamlines of $V_\m$ (expressed with respect to a normalized basis of vectors) in the simpler case $r_0=0$. One can see that \textit{south} of the equator the flow is directed away from the rotation axis, while \textit{north} of the equator the flow is directed towards the axis. Hence the fluid crosses the horizon in opposite directions, so that the acoustic perturbations are all swept outwards in the southern hemisphere and inwards in the northern one. This means that the horizon passes from being a white hole one to being a black hole one.


\section{Acoustic analogues of the BTZ rotating black hole}
\label{sec:btz}
Up to now we mostly bothered with \textit{canonical} acoustic black holes, that is with analogue spacetimes that do not correspond to solutions of the Einstein equations but that show similar causal characteristics, such as horizons and ergosurfaces. Here we will instead focus on the possibility to obtain in relativistic acoustic analogues the causal structure of general-relativistic rotating spacetimes. We start by considering rotating black holes in (2+1)-dimensional gravity.

Gravity in (2+1) dimensions has been studied with the idea to consider conceptual issues in a simpler context than the full (3+1)-dimensional real gravitating systems (\cite{carlip2005conformal}). In many aspects (2+1)-dimensional gravity is very different from from the physical (3+1)-dimensional counterpart, for example it has no propagating degrees of freedom and, with zero cosmological constant, the vacuum solutions of the Einstein field equations are necessarily flat. Moreover, it can be shown that there are no asymptotically flat black hole solutions.

However, when the cosmological constant is negative this is no longer true and it was shown in \cite{btz-original-paper} that (2+1)-dimensional gravity with $\Lam<0$ admits vacuum black hole solutions that in the rotating version share many of the characteristics of the Kerr black hole. These black hole solutions are called \textit{BTZ black holes} from the initials of the authors of \cite{btz-original-paper}.

A rotating BTZ black hole is described by the line element (where $c=G=1$)
\begin{equation}\label{eq:btz-metric}
\begin{split}
ds^2	&=-f^2 dt^2 +f^{-2}dr^2+ r^2\left(d\p+N_\p dt\right)^2\\
	&=-\left(f^2-r^2N_\p^2\right)dt^2-J\ dtd\p - f^{-2}dr^2 +r^2d\p^2,
\end{split}
\end{equation}
where 
\begin{equation}
f\defeq \sqrt{-M+\frac{r^2}{\l^2}+\frac{J^2}{4r^2}}\ ; \quad N_\p\defeq -\frac{J}{2r^2},
\end{equation}
where $M$ and $J$ are the mass and angular momentum.\footnote{Notice that in $(2+1)$-dimensions with $G=1$ masses result to be adimensional.} This line element represents a stationary and axially symmetric metric with Killing vectors $\de_t$ and $\de_\p$.

The spacetime has two horizons at
\begin{equation}
r_\pm^2\defeq\frac{M\l^2}{2}\left\{1\pm\left[1-\left(\frac{J}{M\l}\right)^2\right]^{1/2}\right\},
\end{equation}
and an ergosurface at
\begin{equation}
r_E\defeq\sqrt{r_+^2+r_-^2}=\sqrt{M}\l.
\end{equation}
As for the Kerr black hole we have a limit for the metric parameters to avoid a naked (in this case conical) singularity that is
\begin{equation}\label{eq:btz-non-naked}
M^2\l^2\geq J^2;
\end{equation}
when this is not satisfied there are no horizons.

\subsection{The BTZ metric in non-relativistic acoustic form}
To put the rotating BTZ metric in the shape of a non-relativistic acoustic metric we proceed by changing in (\ref{eq:btz-metric}) the coordinates $t,\p$ to $t',\p'$ and imposing for the term $drd\p'$ to vanish and for the $dr^2$ term to be equal to $1$. This is obtained with the change of coordinates 
\begin{equation}
\begin{dcases}
 dt'= dt-\frac{\sqrt{1-f^2}}{f^2}dr\\
 d\p'= d\p + N_\p\frac{\sqrt{1-f^2}}{f^2}dr.
\end{dcases}
\end{equation}

\begin{figure}[t]
\centering
 \includegraphics[width=0.5\textwidth]{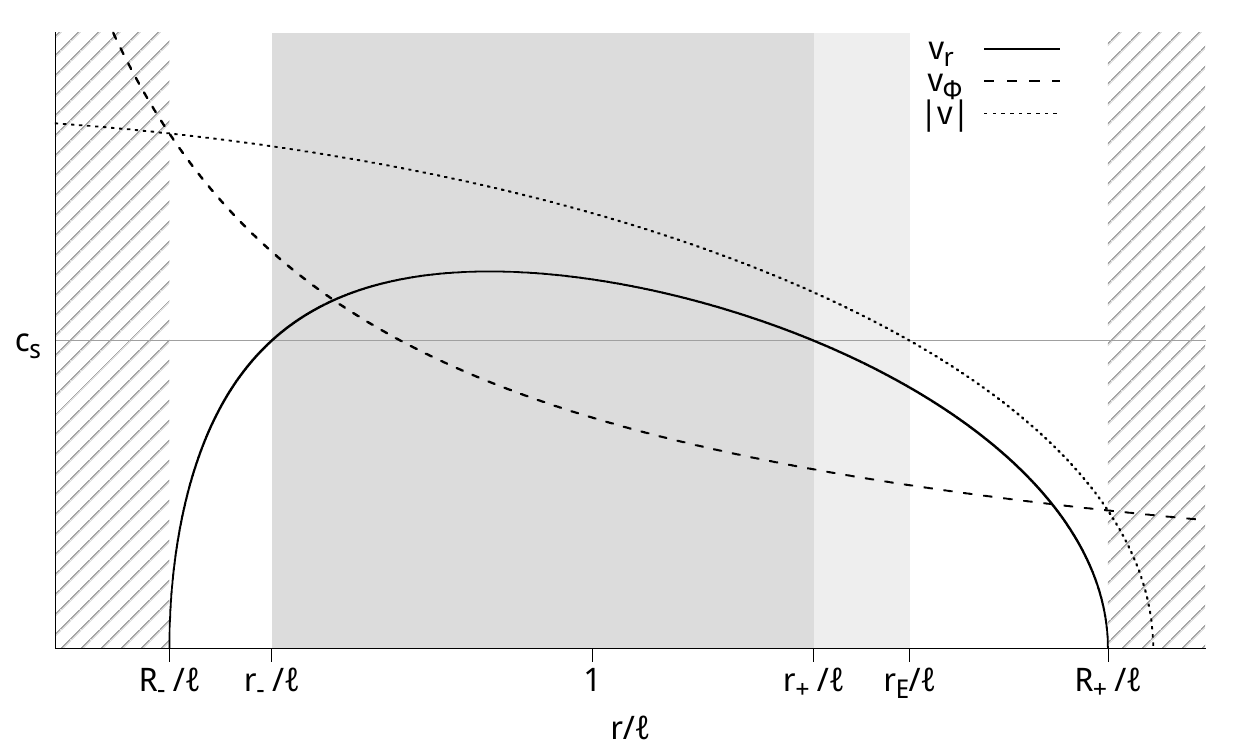}
 \caption{\small\textit{Representation of the velocity of the fluid that gives the rotating BTZ black hole analogue for $M=2$ and $J=1.5\l$. The dashed regions are the ones in which the acoustic simulation breaks down because the radial velocity is ill defined. The dark gray region is the one where the radial flow (solid line) is supersonic, delimited by the horizons that are the points where the radial velocity matches the speed of sound $c_s$. The light gray region instead is the ergoregion, delimited by the ergosurface radius $r_E$.}}
\end{figure}

This brings the line element (\ref{eq:btz-metric}) in the form (also reintroducing $c\to c_s$)
\begin{equation}\label{eq:btz-nonre-line-element}
  \begin{split}
ds^2 =&-\left(f^2-r^2N_\p^2\right)c_s^2dt'^2-2\sqrt{1-f^2}c_sdt'dr\\
	&+J c_sdt'd\p' +dr^2 +r^2d\p'^2 \\
  \end{split}
\end{equation}
that can be compared with the non-relativistic acoustic metric line element (\ref{eq:nonrel-acoustic-metric}) in polar coordinates to see that if we define the velocity (again we indicate with a tilde the components with respect to an orthonormal basis)
\begin{eqnarray}\label{eq:btz-nonrel-vel}
\vc{w}&=&\tld{w}_r\hat{\vc{e}}_r + \tld{w}_\p\hat{\vc{e}}_\p \defeq c_s\sqrt{1-f^2}\ \hat{\vc{e}}_r - rc_sN_\p\hat{\vc{e}}_\p \nonumber\\
	&=& c_s\sqrt{1-f^2}\ \hat{\vc{e}}_r +c_s\frac{J}{2r}\hat{\vc{e}}_\p,
\end{eqnarray}
automatically satisfying the irrotationality condition, we have a non-relativistic acoustic form. 

Notice however that $f^2$ is not always smaller than one for a BTZ black hole, hence the acoustic form of the metric is valid only in the range where this is true, that is where the radial velocity $\tld{w}_r=c_s\sqrt{1-f^2}$ is well defined, and this happens for $R_-\leq r\leq R_+$ with
\begin{equation}\label{eq:btz-range-validity}
R_{\pm}^2\defeq \frac{(1+M)\l^2}{2}\left\{1\pm\left[1-\left(\frac{J}{(1+M)\l}\right)^2\right]^{1/2}\right\}.
\end{equation}
Also $R_-<r_-<r_+<R_+$ so that we can simulate the region of the BTZ spacetime containing the two horizons. 

To have an acoustic analogue including also the ergosurface region we must further restrict the parameters of the BTZ black hole, in particular requiring for the radius of the ergosurface $r_E=\sqrt{M}\l$ to be smaller than the upper radius of the simulated region $R_+$
\begin{equation}
r_E^2<R_+^2\ \Longrightarrow\ M\l^2>J^2.
\end{equation}
This constraint must be added to the condition to avoid a naked singularity (\ref{eq:btz-non-naked}), that assures the presence of the horizons.

\begin{figure}[h]
  \centering
 \includegraphics[width=0.25\textwidth]{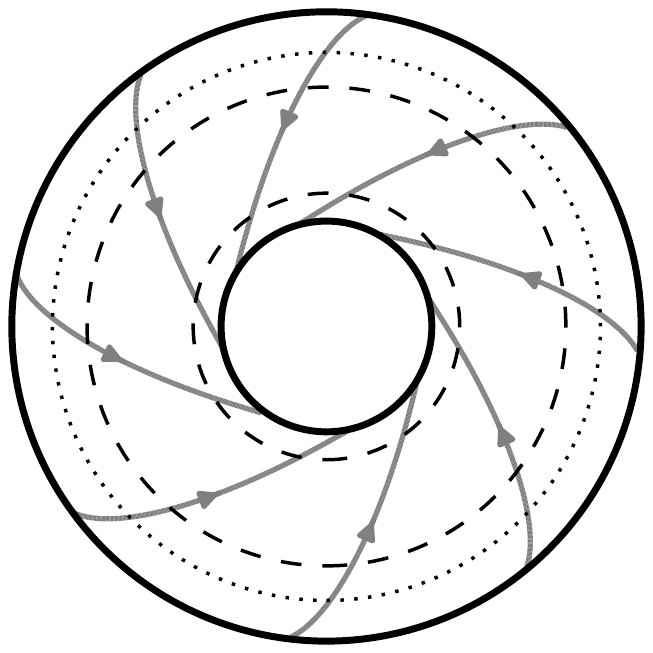}
 \caption{\small\textit{Representation of the streamlines (gray) of the flow that gives the non-relativistic simulation of the BTZ spacetime for $M=2$ and $J=1.7\l$. The solid circles are the limits of the region in which the flow is well defined, and hence have radii $R_{\pm}$. The dashed circles have radii $r_\pm$ and hence are the horizons, while the dotted circle of radius $r_E$ is the ergosurface.}}
\end{figure}

In the region where the analogy holds we still have to impose the continuity equation, that for time-independent matter density is $\del\cdot (\r\vc{w})=0$ and in our case is
\begin{equation}
\frac{1}{r}\pder{(r\rho w_r)}{r}=0.
\end{equation}
Hence we need to fine-tune the density as
\begin{equation}
\rho w_r\propto \frac{1}{r} \Longrightarrow \rho\propto \frac{1}{r\sqrt{1-f^2}}
\end{equation}
to satisfy the continuity equation.

As a check we can compare the radial flow we found here with the one obtained in \cite{dey2016ads} for the simulation in non-relativistic acoustic of the (d+1)-dimensional AdS Schwarzschild black hole. In particular they find for the radial flow in BEC for $d=2$ (equation (40) in the article)
\begin{equation}
u_r=c_s\sqrt{r_0-\frac{r^2}{\l^2}},
\end{equation}
where $r_0$ is a constant. We can compare this with our $\tld{w}_r$, as defined in (\ref{eq:btz-nonrel-vel}), with $J=0$
\begin{equation}
\tld{w}_r=c_s\sqrt{1+M-\frac{r^2}{\l^2}}
\end{equation}
to see that they correspond if we set $r_0=M+1$.

\subsection{The BTZ metric in Gordon form}
In Section \ref{sec:rel-vs-nonrel} we showed how a non-relativistic acoustic metric (that is associated to an irrotational flow) can always be brought in the Gordon form. We now want to apply the transformation to the non-relativistic acoustic form of the BTZ metric we just found. 

We need the transformation (\ref{eq:nonre-rel-transf-inv}) in terms of the non-relativistic velocity components, that with the velocity (\ref{eq:btz-nonrel-vel}) takes the form
\begin{equation}
\begin{split}
dT=dt'&\pm \frac{c_s\sqrt{1-f^2}}{c-\sqrt{c^2-c_s^2(f^2-r^2N_\p^2)}}dr\\
&\pm \frac{c_sr^2 N_\p}{c-\sqrt{c^2-c_s^2(f^2-r^2N_\p^2)}}d\p'.
\end{split}
\end{equation}
Applying this to the line element (\ref{eq:btz-nonre-line-element}) we obtain, in coordinates $(T,r,\p')$, the rotating BTZ metric (with $c\to c_s$) in the Gordon form
\begin{equation}\label{eq:btz-gordon-richiamo}
g_{\mu\nu}=\eta_{\mu\nu}+\xi V_\mu V_\nu
\end{equation}
with unit 4-velocity $V_\m$, in polar coordinates, using (\ref{eq:corresp-nonrel-velocity-def}) and the normalization condition,
{\small
\begin{equation}\label{eq:btz-rel-vel}
V_\mu=\left(\sqrt{1-\frac{c_s^2}{\xi c^2}\left(1-f^2+N_\p^2\right)},\ \frac{c_s}{c}\sqrt{\frac{1-f^2}{\xi}},\ -\frac{c_s}{c}\frac{r^2N_\p}{\sqrt{\xi}}\right).
\end{equation}
}

The range of validity of this form of the metric is the same we discussed for the non-relativistic acoustic form, delimited by the values (\ref{eq:btz-range-validity}) of $r$, and also for the inclusion of the horizons and the ergosurface the discussion is the same. For what concerns the imposition of the continuity equation for a relativistic fluid we have
\begin{equation}
\de_r\left(r n V^r\right)=0\ \iff\ n\propto \frac{1}{r}\frac{c}{c_s}\sqrt{\frac{\xi}{1-f^2}}.
\end{equation}

As we did in the previous section we can again check our resulting flow (\ref{eq:btz-rel-vel}) by comparing its $J\to 0$ limit with the result obtained in \cite{dey2016ads} for the simulation of a (d+1)-dimensional AdS Schwarzschild black hole in a relativistic condensate (equation (33) in the article) specified to $d=2$
\begin{equation}
v_r^2=\frac{c_s^2}{\xi}\left(r_0-\frac{r^2}{\l^2}\right),
\end{equation}
that again coincide provided $r_0=M+1$.


\section{Gordon and Kerr--Schild forms}
\label{sec:kerrschild}
We already said that the Kerr metric cannot be put in a non-relativistic acoustic form because it does not admit conformally flat spatial slices, however in principle one could put it in a Gordon form and obtain a relativistic acoustic analogue of it.

Many different convenient coordinates in which to express the Kerr metric can be chosen, see \cite{visser2007kerr} for a review. One could think to identify the velocity components needed by comparing the transformed Gordon metric (\ref{eq:rel-acoustic-bl}) with a suitable form of the Kerr metric (for example in Boyer--Lindquist coordinates). However that form of the acoustic metric seems not to be suited for any of the most known forms of the Kerr metric. 

Another interesting form of the Kerr metric is the Kerr--Schild form in Kerr--Schild coordinates
\begin{equation}\label{eq:kerr-kerr-schild}
g_{\m\n}=\eta_{\m\n}+\frac{2Mr^3}{r^4+a^2z^2}L_\m L_\n,
\end{equation}
where $L_\m$ is a null 4-vector
\begin{equation}\label{eq:kerr-kerr-schild-null-vector}
L_\m\defeq \left(1,\frac{rx+ay}{r^2+a^2},\frac{ry-ax}{r^2+a^2},\frac{z}{r}\right).
\end{equation}
The Kerr metric in this form is similar to a metric of the Gordon form, but the 4-vector in that case is timelike, since it corresponds to the unit 4-velocity of the medium. A possibility that may come to mind to obtain an acoustic analogue of the causal structure of Kerr is to find a system whose hydrodynamic description has a 4-velocity that is a null 4-vector. Notice also that perturbations on the top of it should move with a sound velocity smaller than the speed of light, otherwise the factor $\xi=1-c_s^2/c^2$ would vanish and the resulting acoustic metric would simply be conformally flat. We do not know if these condition are satisfied by some system, but it would be interesting to investigate.

Sticking to fluids with a timelike 4-velocity field it may seem strange for a metric to admit both Kerr--Schild form and Gordon form, however this is true for the Schwarzschild metric. We discuss this case in detail with the hope to provide some insight towards the simulation of the Kerr metric. 

\subsection{The Schwarzschild case}
In Section \ref{sec:schwarzschild-gordon-form} we showed how the Painlevé--Gullstrand form of the Schwarzschild metric can be cast into a Gordon form. Remember that the associated Gordon metric corresponds to the Schwarzschild spacetime with $c$ substituted by $c_s$, hence if we want to compare it with the Kerr--Schild form we should express $c_s$ in terms of $c$ and $\xi$. Actually it is more instructive to start from the Schwarzschild metric in the usual spherical coordinates and explicit the transformation that brings us to the Gordon form.

Start with the usual Schwarzschild metric
\begin{equation}\label{eq:schwarzschild-line-element}
ds^2_{S}=-\left(1-\frac{2GM}{rc^2}\right)c^2dt^2+\frac{1}{1-\frac{2GM}{rc^2}}dr^2+r^2d\Omega
\end{equation}
and introduce a constant $\xi<1$ and a unit timelike vector $V_\mu$. Now perform the following redefinition of the time coordinate that depends on these quantities
\begin{equation}\label{eq:schw-transf-gordon}
\frac{c}{\sqrt{1-\xi}}dT=\frac{c}{\sqrt{1-\xi}}dt+\frac{\xi V_0V_r}{\xi V_0^2-1}dr,
\end{equation}
which can be rewritten as
\begin{equation}\label{eq:schw-gordon-transf}
c dT=cdt+\sqrt{1-\xi}\frac{\xi V_0V_r}{\xi V_0^2-1}dr.
\end{equation}
The unit timelike vector is chosen as
\begin{equation}\label{eq:schwarzschild-gordon-timelike-vector}
V_r^2=\frac{1-\xi}{\xi}\frac{2GM}{rc^2} \Longrightarrow V_0^2=1+\frac{1-\xi}{\xi}\frac{2GM}{rc^2},
\end{equation}
that is (\ref{eq:schwarzschild-rel-analogue-vel}) with $c^2$ replaced by $c^2/(1-\xi)$, again to obtain the one of the acoustic case when we substitute $c$. With this change of coordinates the Schwarzschild metric takes the form
\begin{equation}
  \begin{split}
  ds_S^2=-\left(1-\xi V_0^2\right)\frac{c^2}{1-\xi}&dT^2 + 2\xi V_0V_r\frac{c}{\sqrt{1-\xi}}dTdr\\
  &+(1+\xi V_r^2)dr^2 +r^2 d\Omega \, ,
\end{split}
\end{equation}
that correctly reduces to a relativistic acoustic metric with the substitution $c\to c_s$. Finally we can reabsorb the $\sqrt{1-\xi}$ factors by redefining the time coordinate as
\begin{equation}
\tld{T}\defeq \frac{T}{\sqrt{1-\xi}},
\end{equation}
so that the Schwarzschild metric assumes the proper Gordon form $g_{\m\n}=\eta_{\m\n}+\xi V_\m V_\n$.

Notice that, besides the acoustic interpretation of this shape of the metric, this is a new form of the Schwarzschild metric, that may be worth investigating on its own.

For what concerns instead the Kerr--Schild form, it can be easily obtained by taking the $a\to 0$ limit of the Kerr--Schild form in Kerr--Schild coordinates of the Kerr metric (\ref{eq:kerr-kerr-schild}), so that the Schwarzschild metric can be written as (reintroducing $G$ and $c$)
\begin{equation}\label{eq:schwarzschild-kerr-schild-form}
g^S_{\m\n}=\eta_{\m\n}+\frac{2GM}{c^2r}L_\mu L_\nu,
\end{equation}
where $L_\mu$ is the null one-form (in spherical coordinates)
\begin{equation}
L_\mu=(1,1,0,0).
\end{equation}

However, this form can also be obtained from the usual expression of the Schwarzschild line element (\ref{eq:schwarzschild-line-element}) by performing the change of coordinates
\begin{equation}\label{eq:schw-kerrschild-transf}
cd\tld{t}=cdt+\frac{\frac{2GM}{rc^2}}{1-\frac{2MG}{rc^2}}dr.
\end{equation}
that is the $\xi\to 0$ limit of the transformation (\ref{eq:schw-transf-gordon}).

We can then combine the two transformations (\ref{eq:schw-kerrschild-transf}) and (\ref{eq:schw-gordon-transf}) to obtain the redefinition of the time coordinate that brings us from the Kerr--Schild form to the Gordon one:
\begin{equation}\label{eq:schwarzschild-kerrschild-gordon-transf}
cdT=cd\tld{t}+\frac{\frac{2GM}{rc^2}-\sqrt{\frac{2GM}{rc^2}}\sqrt{\xi+(1-\xi)\frac{2GM}{rc^2}}}{1-\frac{2GM}{rc^2}}dr.
\end{equation}
Notice that this reduces to an identity for $\xi\to 0$. This reflects the fact that in this limit the timelike 4-vector $\sqrt{\xi}V_\m$ (with $V_\m$ given by (\ref{eq:schwarzschild-gordon-timelike-vector})) reduces to the null vector of the Kerr--Schild form (\ref{eq:schwarzschild-kerr-schild-form}).
Unfortunately, no generalisation of the above transformation for the Kerr geometry has been found up to date. We plan to further explore this issue in future work. 

\vspace{0.15cm}

\section{Conclusions}
In this article we considered analogue gravity in relativistic fluids and relativistic condensates, showing the link between the two. In particular we focused on the relation with the non-relativistic case and on rotating acoustic black holes. Even if these relativistic systems still cannot be obtained in a laboratory, they provide an interesting conceptual tool and may naturally occur in cosmological and astrophysical context.

We showed how every stationary spacetime that can be obtained in non-relativistic acoustic metric with constant speed of sound can be also obtained in the relativistic case by deriving a transformation that maps between the two and we explicitly discussed it for the case of the acoustic analogues of the Schwarzschild spacetime.

We then showed that the relativistic acoustic metric is compatible with the occurrence of rotating acoustic black holes, that is with the presence of horizons and ergoregions, and we presented the relativistic generalization of the vortex (or \textit{draining bathtub}) geometry, showing that this presents some new features with respect to the non-relativistic analogue. We also provided some examples of (3+1)-dimensional configurations.

Eventually, we focused on the acoustic simulation of general-relativistic rotating spacetimes obtaining a fluid configuration that reproduces a region of the causal structure of the rotating BTZ black hole, for both the non-relativistic case and the relativistic one. Finally we commented on the relation between the Kerr--Schild form the Gordon one, analysing in detail the Schwarzschild case, with the aim to lay the foundations of a future generalisation to the Kerr geometry case.

So, in conclusion, the investigations here presented extend the existent results on the general-relativistic objects one can mimic in relativistic analogues and clarifies the relationship with their non-relativistic counterparts. Moreover, the fluid analogies we presented here provide an intuitive picture of rotating spacetimes and might serve as useful toy models for looking at important concepts of general relativity. As a future perspective, we hope that this work will further stimulate the research towards the development of a full Kerr black hole analogue which could provide a very useful test bed for numerical simulations of relevant phenomena involving astrophysical black holes.

\section{Acknowledgments}
 We thank Matt Visser for multiple interesting comments and suggestions on the paper draft and Andrea Trombettoni for illuminating discussions. SL also acknowledges the John Templeton Foundation for the supporting grant \#51876. 

\bibliographystyle{plain}
\bibliography{analogue.bib}
\end{document}